# Mapping Humidity-dependent Mechanical Properties of a Single Cellulose Fibre

Julia Auernhammer[1], Tom Keil[2], Binbin Lin[3], Jan-Lukas Schäfer[4], Bai-Xiang Xu[3], Markus Biesalski[4] and Robert W. Stark[1]*

[1] Technical University of Darmstadt, Institute of Materials Science, Physics of Surfaces, Alarich-Weiss-Str. 16, 64287 Darmstadt, Germany

[2] Technical University of Darmstadt, Institute of Materials Science, Physical Metallurgy, Alarich-Weiss-Str. 2, 64287 Darmstadt, Germany

[3] Technical University of Darmstadt, Institute of Materials Science, Mechanics of Functional Materials, Alarich-Weiss-Str. 16, 64287 Darmstadt, Germany

[4] Technical University of Darmstadt, Department of Chemistry, Macromolecular Chemistry and Paper Chemistry, and Center of Smart Interfaces (CSI), Petersenstraße 22, 64287 Darmstadt, Germany

*corresponding author

stark@pos.tu-darmstadt.de

Phone: 061511621920

Fax: 061511621921

ORCIDs:

| | |
|---|---|
| Julia Auernhammer | 0000-0002-9896-7353 |
| Jan-Lukas Schäfer | 0000-0002-4335-9342 |
| Robert W. Stark | 0000-0001-8678-8449 |
| Markus Biesalski | 0000-0001-6662-0673 |
| Bai-Xiang Xu | 0000-0001-5906-5341 |

Abstract:

Modelling of single cellulose fibres is usually performed by assuming homogenous properties, such as strength and Young's modulus, for the whole fibre. Additionally, the inhomogeneity in size and swelling behaviour along the fibre is often disregarded. For better numerical models, a more detailed characterization of the fibre is required. Herein, we report a method based on atomic force microscopy to map these properties along the fibre. A fibre was mechanically characterized by static colloidal probe AFM measurements along the fibre axis. Thus, the contact stress and strain at each loading point can be extracted. Stress-strain curves can be obtained along the fibre. Additionally, mechanical properties such as adhesion or dissipation can be mapped. The inhomogeneous swelling behaviour was recorded via confocal laser scanning microscopy along the fibre. Scanning electron microscopy measurements revealed the local macroscopic fibril orientation and provided an overview of the fibre topology. By combining these data, regions along the fibre with higher adhesion, dissipation, bending ability and strain or differences in the contact stress when increasing the relative humidity could be identified. This combined approach allows for one to obtain a detailed picture of the mechanical properties of single fibres.

*Keywords:*

*Cellulose, Single Fibre Strength, Scanning Electron Microscopy, Confocal Laser Scanning Microscopy, Atomic Force Microscopy, Colloidal Probe*

Acknowledgements.

The authors would like to thank the Deutsche Forschungsgemeinschaft under grant PAK 962 project numbers 405549611, 405422877, 405422473 and 52300548 for financial support. We thank Lars-Oliver Heim for the modification of the cantilever with the colloidal probe.

Declarations: Nothing to declare.

Abbreviations:

| | |
|---|---|
| AFM | Atomic force microscopy |
| RH | Relative humidity |
| CLSM | Confocal laser scanning microscopy |
| SEM | Scanning electron microscopy |
| SE | Secondary electron |
| ROI | Region of interest |

# Introduction

Cellulose-based papers have promising applications in areas such as electronics, sensor technology, microfluidics, and medicine (Bump et al. 2015; Delaney et al. 2011; Hayes and Feenstra 2003; Liana et al. 2012; Ruettiger et al. 2016). However, before paper can be used as a substrate material for these technologies, an understanding of how the mechanical properties depend on the structure and variations and inhomogeneities of single cellulosic fibres must be improved. Cellulose is a naturally occurring material that is abundant and renewable. It is the most important raw material in the paper-making industry. In pulping and paper making, the flexibility of a single fibre plays an important role. Flexibility is responsible for quality during sheet formation and production of types of papers (Persson et al. 2013). To control this process, characterisation of the processed fibres is important.

Young's modulus is a mechanical parameter that can be used to describe a single fibre. In tensile tests, a Young's modulus between 20 and 80 GPa was measured for single fibres (Groom et al. 2002; Jayne 1959; Lorbach et al. 2014; Page et al. 1977). *Via* an AFM-based three-point bending test, the Young's modulus of a single pine fibre could be quantified to 24.4 GPa (Fernando et al. 2017). Large variations in the Young's moduli of different fibres are due to the different types of fibres, such as earlywood vs latewood, and botanical aspects, such as cell wall thickness or fibre width. Additionally, the micro fibril angle plays an important role in the single fibre strength. A small micro fibril angle was shown to be correlated with a high longitudinal elastic modulus (Müssig 2010).

The parameter of bending stiffness is defined as the Young's modulus multiplied by the second moment of inertia (Samuelsson 1963). The bending stiffness of dry earlywood fibres was tested to be $3.1 \cdot 10^{-5}$ $N^{-1}m^{-2}$, and that of latewood was tested to be $1.5 \cdot 10^{-4}$ $N^{-1}m^{-2}$. When wetting the fibre, the bending stiffness was reduced to $0.9 \cdot 10^{-5}$ $N^{-1}m^{-2}$ for earlywood and to $2.6 \cdot 10^{-5}$ $N^{-1}m^{-2}$ for latewood (Schniewind et al. 1966). Flexibility is another relevant mechanical parameter that characterises single fibres. It is defined as the reciprocal bending stiffness (Yan and Li 2008). Flexibility can be characterised by flowing the fibres between two rotating cylinders (Arlov AP 1958; Forgacs OL 1958) or obtained *via* AFM-based methods. One method is to mount a single fibre over a trench and test the flexibility of the fibre with a colloidal probe attached to the cantilever (Navaranjan et al. 2008). Another approach is to attach the fibre at one end only. At the free end, the cantilever tests the flexibility *via* static force-distance curves. Here, bleached softwood fibres exhibited a flexibility of 4 - $28 \cdot 10^{12}$ $N^{-1}m^{-2}$, and in TMP fibres, the flexibility was one order of magnitude lower (Pettersson et al. 2017). These examples illustrate the drastic effect of humidity on fibre mechanics.

Thus, the influence of humid air or water on the mechanical properties of natural fibres is an important topic to be addressed in the application of cellulose-based materials. The impact of relative humidity (RH) on the elastic modulus, stiffness, or strength has been investigated by various authors (Ganser et al. 2015; Placet et al. 2012; Salmen and Back 1980). With an AFM-based indentation method, it was possible to test the mechanical properties of wet cellulose and estimate the Young's modulus, which was in the kPa range (Hellwig et al. 2018). Additionally, the viscoelastic properties of pulp fibres could be investigated with AFM. Here, the RH was varied from 10 to 75 %, and the elastic modulus and viscosity of the fibre were recorded. The elastic moduli decreased by a factor of ten, and the viscosity decreased by a factor of ten to 20. The fibres exhibited a decrease in the elastic moduli by a factor of 100 after water immersion, and the viscosity decreased by at least three orders of magnitude (Czibula et al. 2019). Further AFM-based colloidal probe measurements on cellulose were performed on gel beads made of cellulose to show the impact on mechanical properties in the wet state (Hellwig et al. 2017). Additionally, the breaking load of a single fibre depending on the RH could be determined, whereas the breaking load decreased with increasing RH (Jajcinovic et al. 2018).

Since cellulose fibres are natural fibres with a hierarchical structure, one must account for the variability in the mechanical parameters within and along the fibre. Earlier, the inhomogeneous swelling behaviour of cellulose fibres was reported (Fidale et al. 2008; Placet et al. 2012). The swelling process of cellulose fibres is determined by the crystallinity, the degree of polymerization, the degree of fibrillation and the pore size (Buschlediller and Zeronian 1992; El Seoud et al. 2008; Fidale et al. 2008; Mantanis et al. 1995). Additionally, variability in cross-sectional areas within the fibre must be considered, as pointed out by (Biswas et al. 2013) and (Chard et al. 2013).

To establish mechanical models of fibres and fibre networks that account for the intra-fibre variability in mechanical parameters, it is essential to locally characterize fibres (Lin 2020). In the following paper, we report an AFM-based method to map the strength, flexibility, and mechanical properties of single cellulose fibres depending on relative humidity and based on different sections of the fibre. Other methods, such as the three-point bending test, are often based on the assumption of a constant cross section along the fibre. However, this assumption might be slightly too simple to yield a good mechanical representation of cellulosic fibres in the dry state and even less in the wet state. To account for the variation in the cross section, we characterised the geometrical and mechanical properties of cellulose fibres along their axis section by section.

First, we observed the swelling behaviour of the local radius of the fibre along its axis *via* confocal laser scanning microscopy. Individual radii, as well as the inhomogeneous swelling behaviour of the radii of the fibre, were considered for each fibre section. Then, we generate static force-distance curves *via* atomic force microscopy with a colloidal probe to test the mechanical properties at every local loading point of the fibre. The local topography was recorded by scanning electron microscopy measurements. A linter fibre, which consists of 95 % cellulose, served as the model system (Mather and Wardman 2015; Young and Rowell 1986), and the results indicate that the mechanical properties of the fibres strongly varied along the fibre axis, particularly in a humid environment.

## Materials and methods

### Materials

Cellulose fibres were manually extracted from a linter paper sheet that was prepared according to DIN 54358 and ISO 5269/2 (Rapid-Köthen process). Further information on the fibre is given in table S1. The extracted fibre was mounted on a 3D printed sample holder, which supplied a fixed trench distance $L$ of 1 mm between the two attachment points. Thus, the fibre under test was freely suspended and fully exposed to the humid environment, which minimized the influences of a substrate or other connecting fibre bonds. The set up was similar to that of other authors (Schmied et al. 2013; Schmied et al. 2012). The glue did not penetrate the fibre, as shown in figure S1.

### Methods

Confocal laser scanning microscopy (CLSM) measurements were conducted to detect the swelling of the paper fibre radius $R_{Fibre}$. With AFM, we could detect the local bending $\delta$, adhesion, and dissipation to complete the mechanical property image with the calculated local occurring contact stress $\sigma$ (unit: N/m$^2$) and strain $\varepsilon$ (unit: %) at each bending point. For the calculation of contact stress $\sigma$, CLSM measurements with differences in $R_{Fibre}$ were used. Regions with notably unique mechanical behaviour or an increase in $R_{Fibre}$ were investigated and related to the local macroscopic fibril orientation on the fibre surface, which was recorded with SEM. In CLSM and AFM measurements, the relative humidity (RH) was varied. For every RH increment, the fibre was measured first with CLSM in a climate chamber and then with AFM in the climate chamber for uniform conditioning and reliable results.

### Confocal Laser Scanning Microscopy

A VK-8710 (Keyence, Osaka, Japan) CLSM instrument was used to investigate the swelling behaviour of the paper fibre. To this end, the fibre was placed into a climate chamber, where the RH was set at 2 %, 40 %, 75 %, and 90 %. The fibre was exposed to the RH for 45 minutes before measuring. As shown in (Carstens et al. 2017; Hubbe et al. 2013), the water progressed into cotton linter test stripes in seconds. Furthermore, (Olejnik 2012) discovered that the free swelling time for whole pulp was 70 minutes. However, (Mantanis et al. 1995) recorded the swelling of single cellulose fibres in water and showed that equilibrium was reached at 45 minutes. Additionally, each experiment at the adjusted RH took at least 30 minutes in total. To ensure consistency between CLSM and AFM measurements and to avoid drift problems later in AFM, we defined 45 minutes of "swelling time" as a feasible compromise to balance effects from instrumental drift and slow swelling processes.

Fibre swelling was analysed using VK analyser software from Keyence (Osaka, Japan). First, the data were noise corrected. Then, cross sections were measured along the fibre to estimate the fibre radius $R_{Fibre}$ of the curvature at each loading point. A representative cross section is shown in figure 1, and the corresponding recorded three-dimensional fibre images are shown in figure S2. For every loading point, such a cross section was drawn. The change in $R_{Fibre}$ with increasing RH was normalized and plotted against the RH. As described in Eq. 9, the radius of the local investigated fibre spot $R_{Fibre}$ is needed to calculate the contact area $A$ (Eq. 12) between the colloidal probe and fibre.

Furthermore, regions of homogeneous swelling behaviour were identified along the fibre diameter. At every loading point, the $R_{Fibre}$ was analysed for every RH. Then, the increase in $R_{Fibre}$ at every loading point with varying RH was noted. The regions of interest (ROIs) were defined as having a similar, normalized increase in $R_{Fibre}$ when varying the RH. The ROIs were then subjected to SEM for further investigation.

As the fibre was fixed at both ends, we wanted to minimize fibre twisting due to swelling. In the evaluation of the radius of the local investigated fibre spot $R_{Fibre}$, we did not observe such a phenomenon.

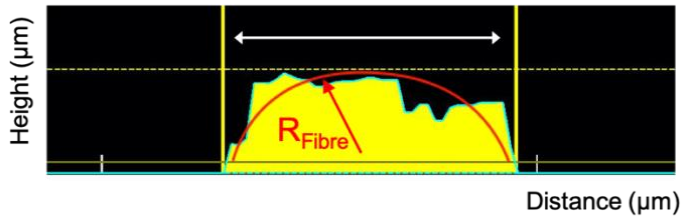

Figure 1: A representative CLSM measurement of the cross section of a loading point of the fibre. The estimated fibre radius is indicated in red.

**Atomic Force Microscopy**

A Dimension ICON (Bruker, Santa Barbara, USA) was used to measure static force-distance curves along the cellulose fibre with a colloidal probe. The cantilever (RTESPA 525, Bruker, Santa Barbara, USA) with a nominal spring constant of 200 N/m was modified with a 50 µm $SiO_2$ colloidal probe (Glass-beads, Kisker Biotech GmbH & Co. KG, Steinfurt, Germany) 50 µm in diameter. The deflection sensitivity of the cantilever (invOLS) was calibrated by performing a force-distance curve on a hard sapphire surface. All AFM experiments were performed in a climate chamber. Hence, it was possible to vary the relative humidity (RH) during the experiments. The chosen RH values were 2 %, 40 %, 75 %, and 90 %. As the RH was adjusted, the fibre was exposed to the environment for 45 minutes before starting the measurements. Force-distance curves were acquired every 5 µm along the freely suspended fibre.

The fibre was fixed at both ends to strain the fibre and prevent twisting of the fibre as the force was applied. Twisting and sideway slippage of the colloidal probe during force curve measurement was investigated in preliminary experiments (data not shown). By trial and error, a colloidal probe diameter of 50 µm turned out to limit slippage on the fibre under test. A step size of 5 µm in the axial direction was set and automatically controlled by AFM software. The lateral position of the colloidal probe was corrected manually based on the camera image. Thus, the scan along the fibre was aligned stepwise along the fibre axis (figure 2). Additionally, fibre bending by the colloidal probe was monitored with the camera image to ensure consistent AFM data. Finally, the recorded force-distance curves were inspected for obvious features of a slipped colloidal probe. In the case of slippage, the probe was repositioned, and the measurement was repeated.

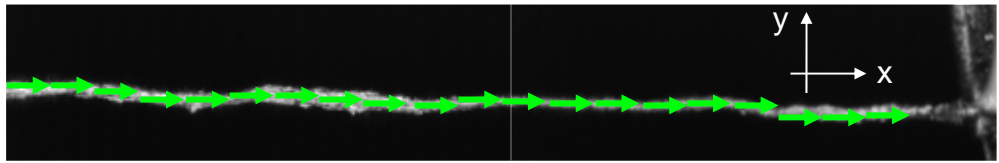

Figure 2: Schematics of the section-wise scanning process along the fibre axis. At every section, the scanning trajectory was realigned to the fibre axis.

Via cantilever calibration, the recorded deflection-piezo position curves were transferred to force-distance curves. A schematic force-distance curve is shown in figure 3a). The peak force setpoint was 2100 nN. The baseline was corrected with a linear fit function to the initial flat region. The contact point was defined with MATLAB (The MathWorks, Inc. Natick, Massachusetts, USA) toolbox code from Bruker. The method first determines the contact point of the curve. The code and its explanation are provided in the Supplementary Information. Mechanical properties such as adhesion and energy dissipation were extracted from the force-distance curve, as illustrated in figure 3a. The adhesion force was identified as the lowest point in the retrace curve (red curve in figure 3a), and the dissipation was defined as the integrated area between the trace and retrace curves (blue and red curves in figure 3a). In figure 3b), a schematic force-bending curve is displayed. To transform the force-distance curves into force-bending curves, the MATLAB code identifies the contact point and sets this point as zero (offset). For the force-distance curve, the offset is in the origin of the curve (cf. position of zero value in figure 3a and 3b). From the data, the fibre bending δ can be read out for various forces from the data to create virtual height maps for these forces, as schematically illustrated in figure 3b. For mapping, we determined the fibre bending for 500 nN, 1000 nN, 1500 nN, and 2000 nN. This allows for one to easily detect a potentially nonlinear local response of the fibre.

Figure S3 shows representative force-bending curves at different RHs. The curves clearly show the nonlinear force-bending relationship. Thus, the bending for different applied forces cannot be easily estimated, as seen in the sketch (figure 3b). Figure S8 illustrates the change in mechanical properties of dry and wetted-and-dried fibres as measured with our method to highlight the added value of plotting topography maps for different forces.

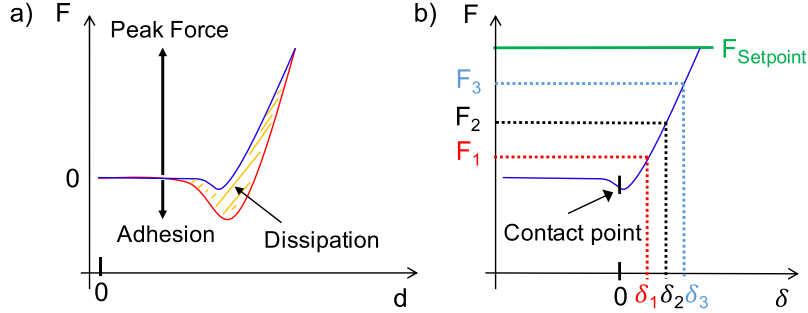

Figure 3: a) Schematic force-distance curve with extracted mechanical properties. b) Force-bending curve with visualization of multiple virtual setpoint forces F.

In contrast to the conventional three-point bending test (figure 4a), where the specimen is bent only at the centre, we performed a test along the fibre axis every 5 µm in a scanning three-point bending test (figure 4c). In conventional three-point bending testing, the second moment of inertia

$$I = \frac{\pi}{4} \cdot (r_2^4 - r_1^4) \qquad (1)$$

is a critical parameter that must be homogenized along the fibre to obtain Young's modulus

$$E = \frac{F \cdot L^3}{192 \cdot \delta \cdot I}. \qquad (2)$$

The calculation of Young's modulus *via* the conventional three-point bending test includes the applied force *F*, the trench length *L*, the bending distance $\delta$, and the second moment of inertia *I*. The second moment of inertia *I* includes the inner and outer radii ($r_1$ and $r_2$) of a schematic circular fibre, as shown in the cross section in figure 4b). Since a cotton linter fibre is a natural fibre, the dimensions along the fibre cannot be modelled using a single value for the cross section. To circumvent this problem, we performed many bending tests and scanned along the specimen to test and determine the mechanical properties. We were able to detect the local bending $\delta$ of the fibre via AFM from the force curves and could, therefore, study the bending behaviour of the fibre at different applied forces. In addition to the local adhesion, dissipation, and bending ability along the fibre axis, we completed the mechanical property imaging with the local stress:

$$\sigma = \frac{F}{A}, \qquad (3)$$

and strain

$$\varepsilon = \frac{\Delta L}{L} \qquad (4)$$

for each loading point.

The strain was obtained by dividing the applied force by the contact area of the colloidal probe and fibre surface. The stress is calculated as the elongation $\Delta L$ divided by the trench length $L$. The estimation of the variables contact area $A$ and elongation $\Delta L$ is described in the following.

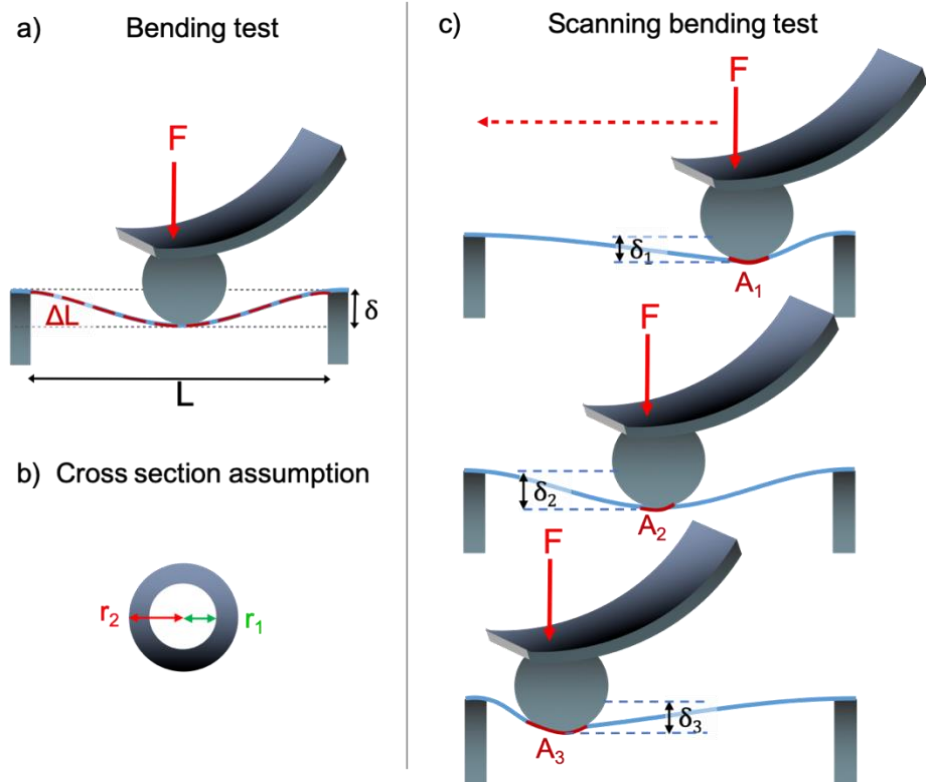

Figure 4: a) Schematic of a conventional three-point bending test. b) Cross section of a fibre and the corresponding equations for a three-point bending test. c) Principle of a scanning three-point bending test.

**Estimation of Contact Area $A$ and Elongation $\Delta L$**

The area $A$ of contact between the colloidal probe and the fibre could not be measured directly; thus, it must be estimated to determine the contact stress. Because the colloidal probe is pressed against the fibre, we assume that rough features at the surface are compressed, which leads to a (nearly) conformal contact between the probe and fibre. To keep the estimation simple, we approximated the area of the elliptical contact area with the area of a rectangle, as shown in figure 5d. The contact mechanics used for the contact length in the x-direction are shown in figure 5a. The model assumes tangents, displayed in dashed red and turquoise, in arc lengths $S_1$ and $S_2$ of the fibre and the colloidal probe with occurring angles $\alpha$ and $\beta$. Via geometric laws, the same angles are also present between the radius $R_{Colloidal}$ and the white dashed middle line. Variables $a$ and $b$ display the segment lengths to the fixed end of the fibre and present loading point. The angles $\alpha$ and $\beta$ are calculated as:

$$\tan(\alpha) = \frac{\delta}{a} \tag{5}$$

and

$$\tan(\beta) = \frac{\delta}{b}. \tag{6}$$

Therefore, the arc lengths $S_1$ and $S_2$ can be defined as

$$S_1 = R_{Colloidal} \cdot \alpha \tag{7}$$

and

$$S_2 = R_{Colloidal} \cdot \beta. \tag{8}$$

The full contact area in the x-direction is then $S_1 + S_2$.

The contact length in the y-direction is determined as the contact mechanics between two spheres, as the curvature of the fibre surface is similar to a sphere, as shown in figures 5b and c (Williams and Dwyer-Joyce 2001). In contact, a resultant contact radius $c$ appears, described as:

$$c = \left(\frac{3 \cdot F \cdot R}{4 \cdot E^*}\right)^{\frac{1}{3}}. \tag{9}$$

The equation includes the applied force $F$, the reduced elastic modulus $E^*$, and the reduced radius of curvature $R$. Both variables are described as:

$$\frac{1}{E^*} = \frac{1-v_{Colloidal}^2}{E_{Colloidal}} + \frac{1-v_{Fibre}^2}{E_{Fibre}} \tag{10}$$

and

$$\frac{1}{R} = \frac{1}{R_{Colloidal}} + \frac{1}{R_{Fibre}}. \tag{11}$$

The elastic modulus of the SiO$_2$ colloidal probe ($E_{Colloidal}$) is 35 GPa (AZoMaterials 2020). The elastic modulus of the fibre was determined as the DMT modulus (Derjaguin et al. 1975) of the fibre surface. The DMT modulus was mapped with a sharp cantilever (Scan Asyst Fluid +, Bruker, Santa Barbara, USA) in PeakForce tapping mode with a force constant of 1.1 N/m, PeakForce Setpoint of 10 nN, and amplitude of 300 nm. Representative DMT modulus maps are shown in figure S4. The DMT modulus was 60 GPa for 2 % RH, 21 GPa for 40 % RH, 2 GPa for 75 % RH and 0.5 GPa for 90 % RH. The Poisson's ratio $v_{Colloidal}$ was defined as 0.17, and the Poisson's ratio $v_{Fibre}$ was defined as 0.3. The radius of the colloidal probe $R_{Colloidal}$ was 50 µm. The local fibre radius $R_{Fibre}$ was recorded via CLSM at every loading point. The full contact area in the y-direction is defined as 2c.

Thus, the resulting contact area $A$ between the colloidal probe and the fibre was

$$A = (S_1 + S_2) \cdot 2c. \tag{12}$$

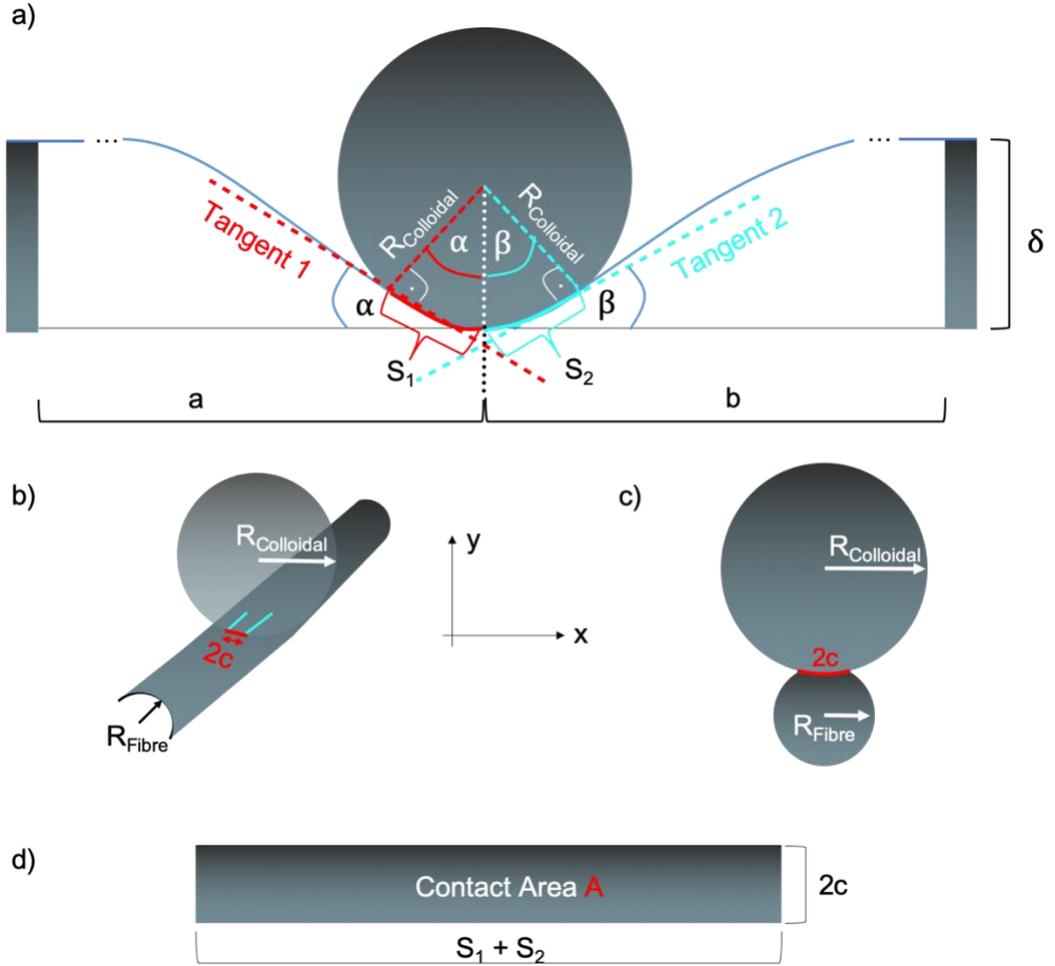

Figure 5: a) Schematics of the contact mechanics of the colloidal probe and the fibre in the x-direction. b) Geometrical dimensions of the fibre segment in contact with the colloidal probe.

To determine the strain along the fibre axis, elongation $\Delta L$ must also be estimated (see Eq.4). Often, the elongated bending length is determined based on a Euler-Bernoulli beam. However, this approach requires homogeneous parameters, and for a fibre, such homogenization is questionable. Thus, we limited bending to small values, and the maximum fibre bending $\delta$ was approximately 2 µm, which is rather small compared to the trench length $L = 1$ mm. Thus, the elongation could be estimated from simple geometrical assumptions with:

$$\Delta L = \text{Tangent 1} + \text{Tangent 2} \qquad (13)$$

as illustrated in figure 6.

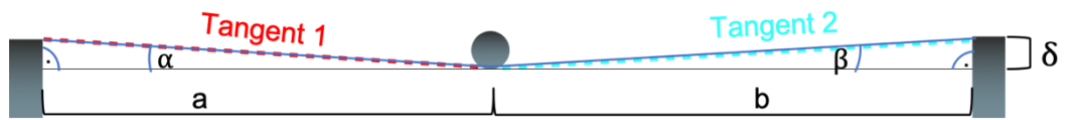

Figure 6: Schematics of the calculation of the elongation $\Delta L$.

**Scanning Electron Microscopy**

The individual fibre topography was analysed using a scanning electron microscope (SEM) (MIRA3, TESCAN, Brno, Czech Republic) in secondary electron (SE) imaging mode at a low acceleration voltage (3 kV).

# Results and Discussion

In the following, we first discuss how the radius of the fibre $R_{Fibre}$ was measured with CLSM and evolved with increasing RH. Then, by combining CLSM and AFM colloidal probe bending measurements, we developed a mechanical image of a single fibre and related the mechanical behaviour and the increase in the fibre radius $R_{Fibre}$ for each ROI.

**Fibre Geometry**

CLSM measurements were used to measure the swelling of the radius of the paper fibre $R_{Fibre}$. These results were used to estimate the contact area $A$ of the colloidal probe and the fibre to determine the contact stress $\sigma$. Figure 7a shows a CLSM image of the fibre. The coloured boxes indicate regions of interest, ROIs, as they were used for CLSM and SEM characterization. The ROIs were defined as regions with similar normalized increases in $R_{Fibre}$ in a humid environment, i.e., the ROIs are defined such that each ROI shows a more or less homogeneous swelling behaviour. In figure 7b, the normalized change in $R_{Fibre}$ of the framed ROIs is displayed over the RH in %. The change in $R_{Fibre}$ was normalized by the value of $R_{Fibre}$ at 2 % RH in each ROI.

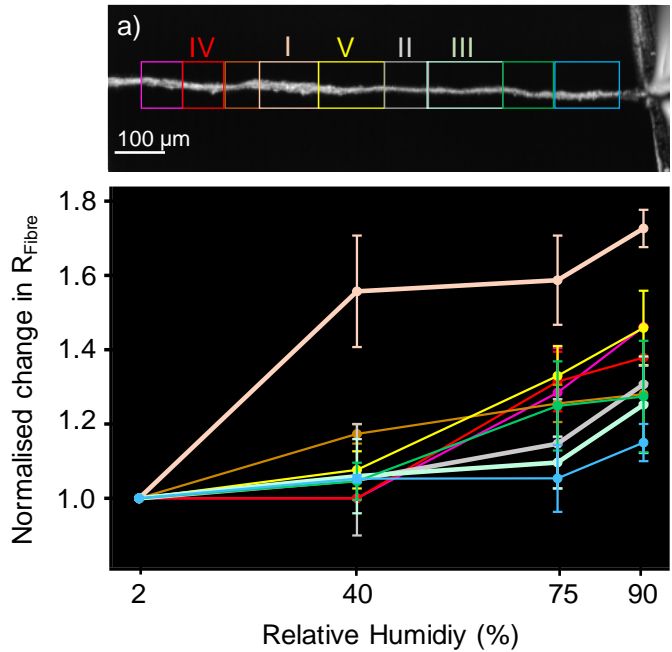

Figure 7: a) CLSM microscopy image of the investigated fibre. The coloured boxed indicate the ROIs. b) The normalised change in $R_{Fibre}/R_{RH=2}$ is displayed over the relative humidity in % of the marked regions of a).

As shown in figure 7, the swelling behaviour of a natural paper fibre is inhomogeneous along the fibre diameter, as indicated with different ROIs. There is general agreement that cellulose fibres with different degrees of polymerization, cellulose contents, crystallinity, and degrees of fibrillation are expected to exhibit different degrees of swelling (Buschlediller and Zeronian 1992; El Seoud et al. 2008; Fidale et al. 2008; Mantanis et al. 1995).

Three ROIs (I, II and III) were investigated in more detail. The corresponding SE images of the discussed ROIs are shown in figure 8. ROI I (light orange) in figure 7a swelled to a normalized change in $R_{Fibre}$ of $1.72 \pm 0.05$ when increasing the RH to 40 %. When increasing the RH to 75 % and 90 %, the radius only slightly increased. The behaviour in ROI II (grey) was different. The swelling was much less until an $R_{Fibre}$ of $1.31 \pm 0.09$ was reached at 90 % RH. In ROI III (green), there was little difference between the radii at 75% and 90 % with a normalised change of $1.25 \pm 0.13$.

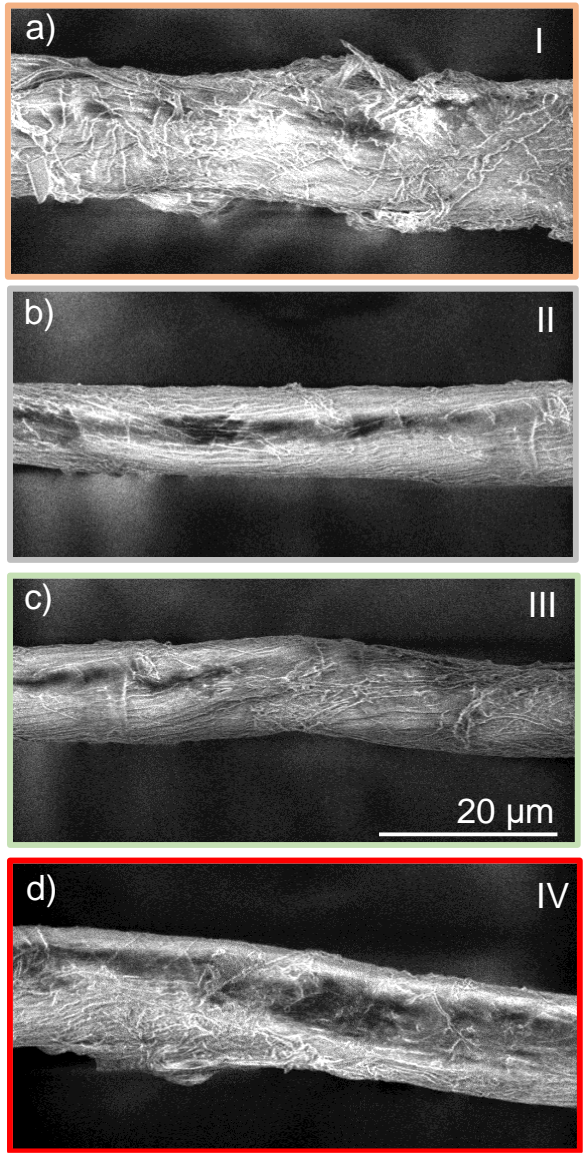

Figure 8: SE images of fibre sections. a) ROI I b) ROI II c) ROI III, d) ROI IV.

As shown in figure 8, the SE images of ROI I exhibited a larger cross section compared to ROIs II and III. In Figure 8a, the SE image revealed macroscopic fibril orientation with little order on the surface in the ROI I. This feature might be an indication for a local internal structure which is different from that in regions with well-ordered macroscopic fibrils. Thus, water uptake leads to distinctively different modes of deformation for both regions. From Figure 8, it is also clear that the fibre cross section varies along the axis. For example, the fibre in ROI II has a smaller cross section than that in ROI I. In ROI II, the macroscopic fibrils are much better aligned than in ROI I, which exists in an ordered, aligned way. Similar observations can be made for ROI III. ROI IV shows a more complex structure of the macrofibrils.

## Colloidal Probe Mapping

Mechanical properties such as local adhesion, energy dissipation, or the bending of the paper fibre were investigated with AFM. Properties such as contact stress and strain were calculated according to Eqs. 3 and 4. The measurements were carried out with a colloidal probe bending the fibre point by point along the axis. Thus, a detailed image of the mechanical behaviour along the fibre could be obtained. Mechanical properties such as adhesion, dissipation, and bending ability were estimated from the force-distance data together with contact stress and strain, as displayed in figure 9. Figure 9a shows the CLSM together with the mechanical properties at 2 % RH, b at 40 % RH, c at 75 % RH, and d at 90 % RH. The shape of the fibre is consistent at all RHs; thus, significant fibre twisting can be excluded.

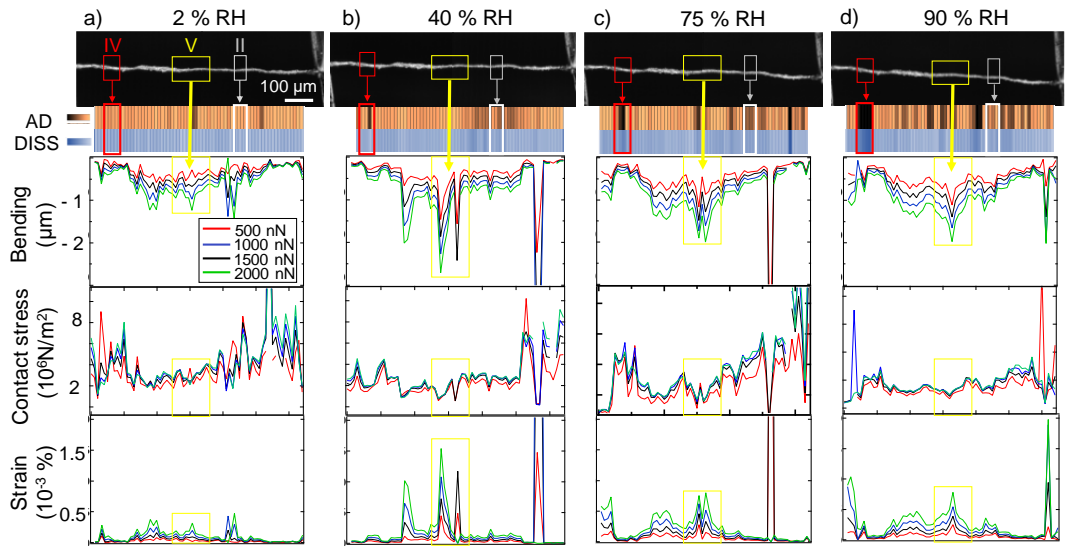

Figure 9: Mechanical characterisation of a cellulose fibre along its axis. From top to bottom: CSLM image of the fibre with the local adhesion and dissipation maps along the fibre beneath. The scale bars are -800 to 0 nN (adhesion) and 1 to 1 x$10^5$ J (dissipation). Beneath: bending behaviour, contact stress and strain along the fibre. The mechanical properties in a) are at 2 % RH, b) at 40 % RH, c) 75 % RH and d) 90 % RH. The discussed ROIs are marked with II, IV, and V.

First, we will discuss the evolution of the adhesion properties with increasing RH. The data in figure 9 clearly show that the adhesion force increases with increasing RH. A comparison of the adhesion maps at varying RHs is provided in figure S5. Due to the hygroscopic properties of cellulose fibre, water molecules can diffuse inside the cellulose network and wet the fibre (Cabrera et al. 2011; Gumuskaya et al. 2003; John and Thomas 2008; Lindman et al. 2010). The wet fibre can promote capillary condensation, which gives rise to strong capillary forces between the colloidal probe and fibre surface. When the probe is retracted, large forces are required to break the capillary neck (Feiler et al. 2005; Fukunishi and Mori 2006; Tejedor et al. 2017). Since capillary forces are hysteretic in nature, this also implies an increase in energy dissipation. Thus, adhesion and dissipation increase with increasing RH. In general, two trends in adhesion changes with increasing RH could be observed: (i) a strong increase in adhesion with increasing RH, as observed in ROI IV, and (ii) a minor increase in adhesion, as in ROI II. The corresponding topographies with their macroscopic fibril orientation in both ROIs are shown in the SE images in figure 8. Figure 8b) displays the topography of ROI II, which exhibits an ordered macroscopic fibril orientation along the fibre. Figure 8d reveals more macroscopic fibrils oriented on the surface in ROI IV than in ROI II. Thus, the data suggests that the number of macroscopic fibrils is correlated with adhesion. One must speculate about the reason for this correlation. Most likely, areas with a large number of disordered macroscopic fibrils have a different internal structure than regions without macroscopic fibrils. Thus, regions with disordered fibrils on the surface show a stronger response to water uptake. The most likely reasons for this behaviour are differences in the local mechanical properties or differences in the local water uptake.

From the bending data in figure 9, the local bending increased when increasing the RH from 2 % RH to 90 %. Local bending lines are given for selected forces in figure S6. Additionally, mechanical effects (clamping) play a role: the bending of the fibre is reduced at both fixed ends and is increased in the middle of the fibre. Here, the reduced bending at both ends is due to attachment to the sample holder.
An interesting observation is marked in the ROI V in figure 9. At 2 % RH, the bending behaviour looks normal with increased bending ability in the middle of the fibre. When increasing the RH to 40 %, ROI V exhibited two significant notches at positions where the fibre was particularly soft. The soft regions (notches) broadened and were no longer distinctive at 75 % RH. When increasing the relative humidity further to 90 % RH, only one notch (=soft region) could be distinguished. A possible interpretation is that by increasing the RH, more water molecules intrude into the cellulose network, weakening the internal structure (Cabrera et al. 2011; Gumuskaya et al. 2003; John and Thomas 2008), which reduces the stability of the fibre. In addition, the water molecules act as lubricants between the fibrils, which leads to the ability of higher bending at higher RH as the fibrils can slip along each other. As indicated in ROI IV in figure 9, some areas exhibit distinctive regions with strong local deformation (notches in the plot). Thus, one might speculate that at 40 % RH, "wet spots" develop, which weaken the fibre in distinctive regions. Increasing the RH further to 90 % RH, more water molecules diffuse in the other parts of the fibre, which leads to more uniform deformation along the fibre axis.
To verify that the bending behaviour displays the actual bending behaviour with its features, we performed a second set of experiments where we characterized a fibre for two RH cycles (see S7 and S8). Additionally, in this verification experiment, the bending ability of the fibre was decreased at both fixed ends due to clamping. In other regions of the fibre, "soft spots" again developed.
Furthermore, figure S7 shows that the increase in the normalised $R_{Fibre}$ over the RH for every ROI changes and differs within the two different cycles. In figure S8, changes between the two cycles of varying the RH in adhesion and dissipation properties as well as in the bending behaviour are observable. In the second cycle of changing the RH, the setpoint of 2100 nN could not be reached in most spots.

Thus, it can be suggested that with wetting of the fibre or exposing the fibre to RH, fibre strength is lost.

This fact also highlights the value of our method. By reading out the "virtual setpoint forces" beneath the setpoint, the mechanical properties of the fibre can be displayed at different setpoint forces at a uniform fibre state. Nonlinearities or peculiar effects, such as local softening, can be detected in a straightforward manner.

In the contact stress graphs in figure 9, it is evident that the contact stress was higher at both ends of the fibre. This is due to the attachment at the fixed ends and should not be interpreted as a higher strength of the fibre. The contact stress at both fixed ends decreases as the RH is increased. Additionally, here, the increased amount of water molecules inside the cellulose network destroys H-bonds and acts as a lubricant between the fibrils, which leads to fibre softening. The strain displayed in figure 9 is linked to the bending properties of the fibre. Thus, the strain behaviour follows the same trend as the bending behaviour and is assumed to be interpreted identically.

Furthermore, the applied method makes it possible to establish stress-strain diagrams for each loading point. The stress-strain diagrams for ROIs I, II, and III are displayed in figure 10. The occurring errors are estimated to be 25 % in strain due to the geometrical assumptions and 30 % in the contact area, as the real shape of the contact area could differ from the estimated contact area.

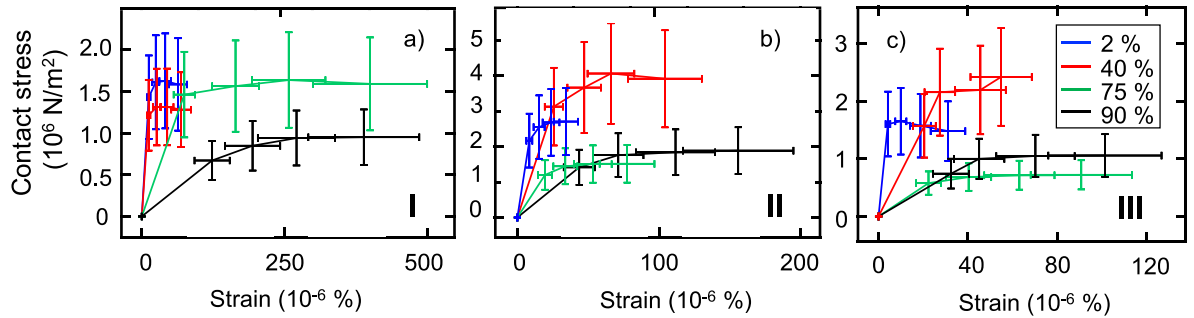

Figure 10: Stress-strain diagrams of ROI I in a) of ROI II in b) and of ROI III in c).

The stress-strain diagram of ROI I is shown in figure 10a. It is evident that the slopes in the linear-elastic area decrease with increasing RH. At 2 % and 40 % RH, linear-elastic and non-linear elastic areas are monitored but no elastic-plastic areas are monitored. At 75 % and 90 % RH, the non-linear elastic area is distinctively larger than that at 2 % and 40 % RH. However, there is no clear indication for an elastic-plastic area. Therefore, we suggest that there is no plastic deformation in the fibre structure in ROI I.

Figure 10b shows the stress-strain diagram of ROI II. The slopes in the linear-elastic area also decrease with increasing RH. The non-linear-elastic area increases with increasing RH. Additionally, no clear indication for an elastic-plastic area could be observed.

ROI III, shown in figure 10c, exhibited the same behaviour as ROIs I and II. The slopes in the linear-elastic area decreased with increasing RH, and the non-linear-elastic regime increased with increasing RH. Additionally, no clear indication for an elastic-plastic regime can be found within the error bars.
These observations strengthen the assumption that the fibre could fully recover its structure between the static colloidal probe experiments, and our method with applying multiple setpoints could additionally preserve the fibre.

With Hooke's law, the Young's modulus can be determined in the linear elastic area. The calculated Young's moduli for every RH in the three ROIs are shown in table 1.

Table 1: Corresponding Young's moduli of ROIs I, II, and III of figure 10.

| RH (%) | E (GPa) ROI I | E (GPa) ROI II | E (GPa) ROI III |
|---|---|---|---|
| 2 | 25.4 ± 7.6 | 95.4 ± 28.6 | 94.4 ± 28.3 |
| 40 | 25.0 ± 7.6 | 71.4 ± 21.4 | 52.5 ± 15.7 |
| 75 | 14.3 ± 4.3 | 46.1 ± 13.8 | 18.0 ± 5.4 |
| 90 | 4.0 ± 1.2 | 13.1 ± 3.9 | 17.6 ± 5.2 |

The Young's moduli in ROI I exhibited lower values than those in ROIs II and III (Table I). Typical values for the Young's modulus for plant-based cellulose are 20-80 GPa but can also amount to 138 GPa for highly crystalline cellulose (Cabrera et al. 2011; Eichhorn and Young 2001; Groom et al. 2002; Jayne 1959; Kolln et al. 2005; Lorbach et al. 2014; Nishino et al. 1995; Page et al. 1977). Furthermore, the Young's modulus is dependent on geometrical aspects, such as cell wall thickness or fibre width. In figure 8, the corresponding SE images of the ROIs are shown.
ROI I exhibits a larger diameter and unordered macroscopic fibril orientation on the fibre surface. In ROIs II and III, the SE images in figure 8 show smaller diameters and an ordered macroscopic fibril orientation on the surface. This suggests that the order and arrangement of the microscopic fibrils could be an indicator of the mechanical strength of the fibre. Additionally, with respect to the mechanics, it seems reasonable that unordered macroscopic fibrils are indicative of a reduced local stiffness of the fibre.
All values of the Young's moduli in table 1 decreased with increasing RH. This is in good agreement with other reports in the literature (Czibula et al. 2019; Ganser et al. 2015; Hellwig et al. 2018; Placet et al. 2012; Salmen and Back 1980). The values of table 1 were normalised and plotted in figure 11. From figure 11, it is clear that all normalised Young's moduli from the different ROIs decreased to the same range of values at 90 % RH. In ROI I, the normalised Young's moduli decreased to 0.15, 0.14 in ROI II, and 0.18 in ROI III. This means that we observed a decrease in the Young's moduli of 6-7 times the initial value at 2 % RH. As we did not immerse the fibre in water to cause a drop of 10-100 times the Young's moduli and obtain values in the kPa range as (Hellwig et al. 2018), we think a decrease of 6-7 times at 90 % RH aligns well with (Czibula et al. 2019).

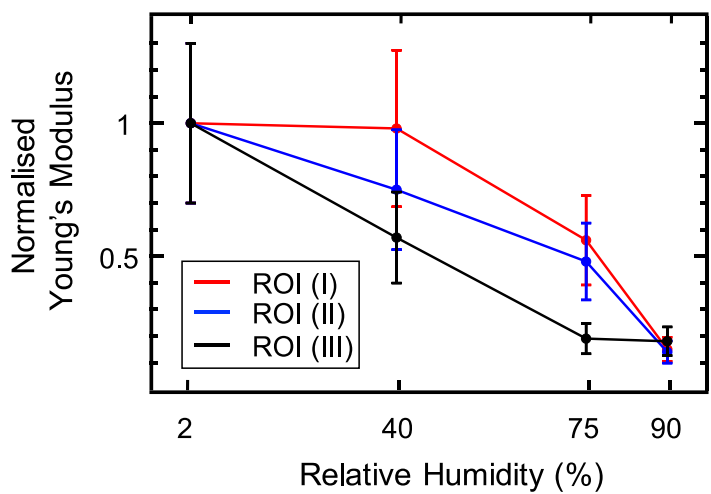

Figure 11: Young's modulus (normalised to the modulus at RH = 2 %) versus the RH for ROIs I-III.

## Conclusions

We measured the mechanical properties of a free hanging cellulose fibre (clamped at both ends) with colloidal probe AFM along the fibre axis. The proposed method of scanning along the fibre axis provides a more detailed picture of the mechanical behaviour than conventional three-point bending tests, where only one measurement is performed. To demonstrate the potential of this approach, the mechanical properties of a single cellulose fibre were mapped for varying relative humidity. The colloidal probe data were analysed in correlation with data from confocal laser scanning microscopy and scanning electron microscopy. The combination of these methods allowed for insights into the interdependence of swelling, bending ability, contact stress, and the stress-strain curves depending on the macroscopic fibril structure and orientation.

With the help of CLSM, the swelling of the fibre radius of different parts of the cellulose fibre could be identified. Regions with a lower order in the macroscopic fibril orientation on the surface showed immediate deformation, and regions with a more ordered orientation exhibited steady volume increase.

With the help of force-bending curves obtained with a colloidal probe AFM, it was possible to create a detailed mechanical picture of the paper fibre. Mechanical properties such as adhesion and dissipation, bending ability, contact stress, and strain could be identified. In general, the bending ability increased as RH increased. Additionally, "soft spots" could be identified at an intermediate RH of 40 %. Increasing the RH, the soft areas became less distinctive when the entire fibre became softer. This implies that at intermediate RH, individual soft spots can occur in a cellulose fibre weakening the fibre locally, whereas the entire fibre softens at elevated RH. By analysing the AFM force curves, it was possible to generate stress-strain curves for local points on the fibre. The stress-strain curves were dependent on the macroscopic fibril orientation: regions with an ordered macroscopic fibril orientation on the surface appeared to be stronger than those with less order. From the stress-strain diagrams, the local Young's moduli could be identified, and the decrease in values with increasing RH was investigated. The decrease in Young's moduli from 2 % RH to 90 % RH was approximately 6-7 times. The measured local Young's moduli illustrate the variations within a cellulose fibre. Thus, our approach highlights the difficulties in the three-point bending test, where the fibre is assumed to be homogeneous along its axis. Depending on the loading point, the corresponding diameter and macroscopic fibril orientation, we found different Young's moduli.

In future work, it will be interesting to address the internal fibre and fibril structure by microtomography or *via* fluorescent dyes in confocal fluorescence microscopy. Additionally, a better model for the contact area of a colloidal probe and the fibre will be relevant to further develop the method. Thus, better statistics of local mechanical parameters will help to develop computer models of fibres or fibre networks that account for the inhomogeneity of the fibres.